# High-Fidelity Universal Quantum Gate Compilation for Non-semisimple Ising Anyons via Genetic Algorithm-Optimized Solovay-Kitaev Decomposition

Jiangwei Long [1], Zihui Liu [1], Yizhi Li [2], Jianxin Zhong [3] and Lijun Meng [1, †]

[1] School of Physics and Optoelectronics, Xiangtan University, Xiangtan 411105, Hunan, People's Republic of China

[2] School of Physics and Electronic Science, Hunan Institute of Science and Technology, Yueyang 414006, People's Republic of China

[3] Center for Quantum Science and Technology, Department of Physics, Shanghai University, Shanghai 200444, People's Republic of China

Non-semisimple Ising anyons extend the fusion space of conventional Ising models and offer a new route toward universal topological quantum computation (TQC). However, how to construct a high-fidelity universal quantum gate set that meets fault-tolerant thresholds via braiding operations of non-semisimple Ising anyons remains unexplored. By employing a Genetic Algorithm-enhanced Solovay-Kitaev Algorithm (GA-enhanced SKA), we achieve high-fidelity approximations of standard single-qubit gates (Hadamard *H*-gate and phase *T*-gate) with a recursion level of just three, meeting the fidelity requirements for fault-tolerant quantum computation. Our numerical results demonstrate that for the critical parameter range $\alpha \in [2.001, 2.022]$, a few braiding operations can approximate the local equivalence class [CNOT] with high precision. In contrast to prior theoretical works that established the universality of non-semisimple Ising anyons, this work provides the first complete, executable, and parameter-optimized compilation scheme for high-fidelity universal gates. It quantifies the relationship between the key physical parameter α and gate fidelity, and it delivers direct numerical evidence for the universality of non-semisimple Ising anyons.

## 1 Introduction

Quantum computing has demonstrated significant advantages over classical computing in solving specific problems [1]. However, quantum systems remain highly susceptible to environmental noise, leading to decoherence issues. Kitaev's proposal of utilizing anyons for topological quantum computation (TQC) offers a promising direction to address these challenges [2]. The primary advantage of this approach lies in its utilization of topological properties to encode quantum information globally, thereby significantly enhancing the system's inherent resilience to local perturbations [3]. The implementation of TQC relies on the braiding [4], measurement [5], and fusion operations of non-Abelian anyons.

The $SU(2)_k$ anyon model describes quasiparticle excitations within topological phases with its mathematical foundation based on the *k*-level unitary representations of the SU(2) group. These models characterize the behavior of anyons exhibiting non-trivial exchange statistics in two-dimensional systems [6]. Theoretical [7,8] and numerical [9] studies have established that for $SU(2)_k$ models with $k \geq 3$ and $k \neq 4$, universal quantum computation can be achieved through braiding operations alone. The $k$=3 Fibonacci anyon model represents the simplest non-Abelian anyon model capable of universal quantum computation solely via braiding operations. Extensive research has been conducted on theoretical constructions of quantum gates using Fibonacci anyons, including one-qubit [10], two-qubit [11], three-qubit [12], and $N$-qubit [13] gates. Similarly, the $k$ = 4 metaplectic anyon model has been proven theoretically universal when supplemented by fusion and measurement operations [14-17]. Recent advances have Introduced cabling concepts from knot theory to achieve low-leakage entangling gates [18].

---

† Corresponding author. E-mail: ljmeng@xtu.edu.cn

The physical realization of both Fibonacci and metaplectic anyons remains experimentally challenging, particularly for $SU(2)_k$ models with $k > 4$. The $k = 2$ case corresponds to Ising anyons, whose proposed physical embodiment involves Majorana fermions. These are considered the most promising candidates for non-Abelian anyons realization, potentially existing in fractional quantum Hall systems [19] and topological superconductors [20]. Recent research has extensively explored quantum gate construction using Majorana fermions [21-23]. However, the standard Ising anyon models faces a fundamental limitation: its inability to achieve universal quantum computation through braiding operations alone, as the $T$-gate ($\pi/8$ phase gate) cannot be implemented [24]. Consequently, supplementary operations are required to establish universality [25].

Recent work by Filippo Iulianelli et al. introduced a modification to the conventional Ising anyon model based on non-semisimple topological quantum field theory. This modification incorporates a *neglecton* $\alpha$ (a new anyon type indexed by non-half-integer real numbers, $\alpha \in (2, 3)$) with traditional quantum trace zero, which remains stationary during the braiding processes [26]. Within this revised model, braiding operations can generate a dense cover of the SU(2) group. Furthermore, by leveraging Reichardt's algorithm, arbitrary entangled gates with arbitrarily low leakage error can be realized [27]. The Hilbert space of the non-semisimple Ising anyon model is closely related to that of the open XXZ spin chain with modified boundary conditions [28], suggesting that fractional quantum Hall states hosting neglectons may be realized within existing experimental paradigms by engineering novel boundary conditions. This work leverages the elementary braiding matrices (EBMs) for both one- and two-qubit operations within the non-semisimple Ising anyon model to construct a high-fidelity universal gate set {$H$-gate, $T$-gate, CNOT-gate}. Here, the high-fidelity of $H$- and $T$-gates are obtained via a Genetic Algorithm-enhanced Solovay-Kitaev Algorithm (GA-enhanced SKA), whereas the CNOT-gate is implemented using only a few braiding operations natively. Our methodology provides a novel approach for achieving universal quantum computation with this model, addressing previous limitations through optimized compilation techniques.

The paper is structured as follows: Section 2 introduces the non-semisimple Ising anyon model and the GA technique. Section 3 presents the numerical results of our gate compilation. Section 4 provides a concluding summary.

## 2 Theoretical Framerwork and Computational Methods

Compared to the conventional $SU(2)_2$ model, the Ising anyon model based on a non-semisimple topological quantum field theory incorporates additional particles with topological spin-2 and spin-3/2 (denoted as $P_2$ and $S_{3/2}$), along with the *neglecton* α, all exhibiting a quantum trace of zero. The fusion rules for this modified Ising anyon model are as follows[26,29]:

$$\begin{aligned} &V\otimes I=V,\ \sigma\otimes\sigma=I\oplus\psi,\\ &\sigma\otimes\psi=\sigma\oplus S_{3/2},\ \sigma\otimes S_{3/2}=P_2,\ \psi\otimes\psi=I\oplus P_2,\\ &\alpha\otimes\sigma=(\alpha+1)\oplus(\alpha-1),\quad \alpha\otimes\psi=(\alpha+2)\oplus\alpha\oplus(\alpha-2), \end{aligned} \tag{1}$$

where the symbol ⊗ denotes fusion of two anyons, ⊕ indicates possible fusion outcomes (i.e., the types of anyons that may result), σ represents the Ising anyon, $\psi$ denotes the fermion, and $I$ stands for the vacuum. By removing $P_2$ and $S_{3/2}$ from these fusion rules recovers the fusion rules of the conventional Ising anyon model.

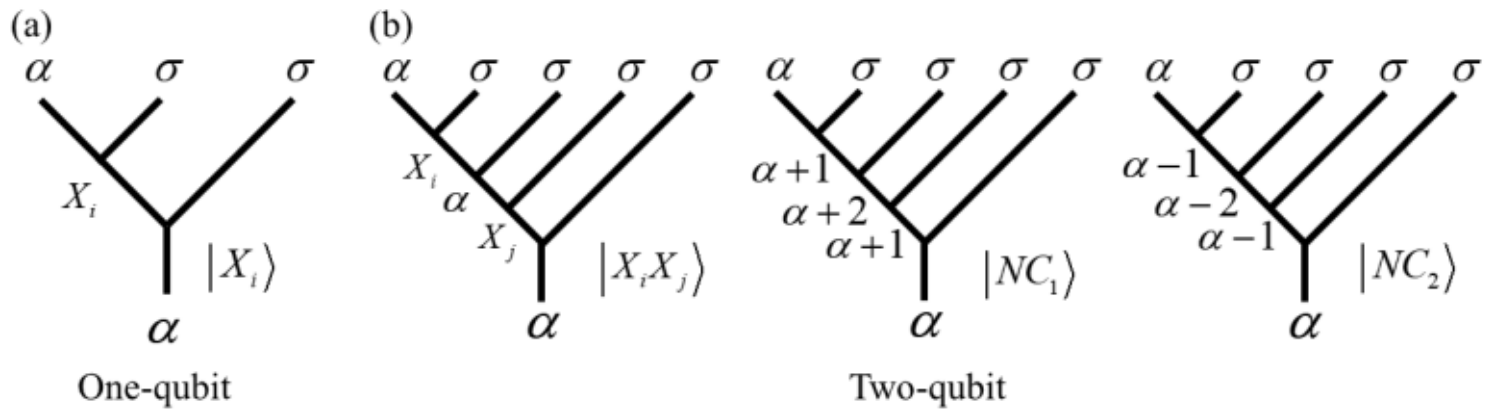

**Fig. 1**: (a) One-qubit is composed of a neglecton α and two Ising anyons σ. (b) Two-qubit system is composed of a neglecton and four Ising anyons σ. The first state encodes the computational state, while the second and third states correspond to non-computational states. This encoding scheme based on non-semisimple Ising anyons originates from Ref. [26,29].

We closely follow the Ref. [26,29] in this section. As shown in Fig. 1(a), a single qubit comprises three anyons: one neglecton α and two Ising anyons σ. According to the fusion rules in Eq. (1), fusing α with the first σ yields two possible intermediate outcomes (α+1 and α-1). Subsequent fusion with the second σ returns the total charge to the final state α. This fusion process encodes the qubit, where the basis states $|0\rangle$ and $|1\rangle$ correspond to the intermediate fusion states $|\alpha+1\rangle$ and $|\alpha-1\rangle$, respectively. The EBMs for one-qubit operations within the computational basis $\{|0\rangle,|1\rangle\}$ are given as follows:

$$\left(b_1^{(3)}\right)^2=\begin{pmatrix} q^{3+\alpha} & 0\\ 0 & q^{3-\alpha}\end{pmatrix},\quad b_2^{(3)}=q^{\frac{1}{2}}\begin{pmatrix} \dfrac{1+q^2}{1-q^{2\alpha}} & q^{-1}\dfrac{\sqrt{B_{\alpha+1}}}{\sqrt{B_{\alpha-1}}}\\ q^{-1}\dfrac{\sqrt{B_{\alpha+1}}}{\sqrt{B_{\alpha-1}}} & \dfrac{1+q^2}{1-q^{-2\alpha}}\end{pmatrix}, \tag{2}$$

$$B_{\alpha+1}=\frac{\sqrt{2}}{-1+\cot\dfrac{\pi(\alpha+1)}{4}},\quad B_{\alpha-1}=\frac{\sqrt{2}}{-1+\cot\dfrac{\pi\alpha}{4}},$$

where $q$ is set to be an eighth root of unity $e^{i\pi/4}$. The symbol $b_i^{(3)}$ denotes the braiding between the $i$-th and ($i$+1)-th anyons, with the superscript (3) indicating this braiding matrix corresponds to a three-anyon one-qubit system. The term $\left(b_1^{(3)}\right)^2$ indicates that the first anyon (the neglector α) and second anyon (Ising anyon σ) must be braided twice consecutively, necessary due to their different anyon types requiring two exchanges to restore the original configuration.

The two-qubit system, as shown in Fig. 1(b), extends the one-qubit configuration by simply adding two σ anyons. By fixing the fusion outcome of the second intermediate state as α, the first and third intermediate states $\left|X_i X_j\right\rangle$ ($\left|\alpha+1,\alpha+1\right\rangle$, $\left|\alpha+1,\alpha-1\right\rangle$, $\left|\alpha-1,\alpha+1\right\rangle$, $\left|\alpha-1,\alpha-1\right\rangle$) encode the computational basis states ($|00\rangle,|01\rangle,|10\rangle,|11\rangle$). According to the fusion rules in Eq. (1), sequential fusion of initial α with four σ anyons, returns to overall α, generating two non-computational states $\left|NC_1\right\rangle$ and $\left|NC_2\right\rangle$ alongside the four computational basis states, as illustrated on the right side of Fig. 1(b). The EBMs for two-qubit operations within basis $\left\{|00\rangle,|01\rangle,|10\rangle,|11\rangle,|NC_1\rangle,|NC_2\rangle\right\}$ are:

$$\begin{aligned}
&\left(b_1^{(5)}\right)^2=\left(b_1^{(3)}\right)^2\otimes I_2\oplus\left(b_1^{(3)}\right)^2,\\
&b_2^{(5)}=b_2^{(3)}\otimes I_2\oplus\left(q^{\frac{1}{2}}I_2\right),\\
&b_4^{(5)}=I_2\otimes b_2^{(3)}\oplus\left(q^{\frac{1}{2}}I_2\right).
\end{aligned}\tag{3}$$

where $I_2$ denotes the two-dimensional identity matrix. The superscript in $b_i^{(5)}$ indicates EBM for five-anyon two-qubit system, distinguishing it from the EBM of the one-qubit system. The subscript $i$ represents the braiding of the $i$-th and ($i$+1)-th anyons. Due to its non-trivial structure, the EBM $b_3^{(5)}$ does not decompose into a simple direct product or direct sum form; its explicit form is provided in SUPPLEMENTAL SECTION I. We have verified the correctness of the EBMs derived from Ref. [26] ; the detailed calculations of these EBMs are provided in SUPPLEMENTAL SECTION I.

A universal quantum computation gate set [30] comprises :

$$H=\frac{1}{\sqrt{2}}\begin{pmatrix}1&1\\1&-1\end{pmatrix},\quad T=\begin{pmatrix}1&0\\0&e^{i\pi/4}\end{pmatrix},\quad \mathrm{CNOT}=\begin{pmatrix}1&0&0&0\\0&1&0&0\\0&0&0&1\\0&0&1&0\end{pmatrix}.\tag{4}$$

We focus on the standard {$H$, $T$, CNOT} universal gate set due to its widespread adoption in quantum computing, but our compilation framework is general and can be extended to other universal gate sets as needed.

Standard *H*-/*T*-gates cannot be constructed using few braiding operations, making Brute-Force search (BF search) infeasible. This necessitates quantum compilation approaches analogous to those for Fibonacci anyons, where extended braid sequences approximate target one-qubit gates [10]. Various methods have been developed for Fibonacci anyon gate compilation:, including algebraic techniques [31], generic approaches [32], reinforcement learning [33], Monte Carlo-enhanced Solovay-Kitaev algorithms [34], and GA-enhanced SKA [35]. We employ GA-enhanced SKA to construct standard *H*-/*T*-gates from non-semisimple Ising anyon EBMs, selected for its low computational cost and high-precision gate synthesis capability. A brief description of this method is provided below.

Topological quantum compilation refers to the process of constructing standard one-qubit gates through the braiding operations of anyons. This involves systematically combining the EBMs of a specific anyon model to form a braidword of length $l$, where $l$ corresponds to the number of EBMs used, in order to achieve a high-fidelity approximation of the target gate. A metric is required to quantify the similarity between the constructed braidword and the ideal one-qubit gate. The global phase-invariant distance serves as an excellent choice for this purpose, as it inherently disregards the global phase, which is physically irrelevant in quantum computation [3]. The metric is defined as follows:

$$d\left(U_0,U\right)=\sqrt{1-\frac{\left|Tr\left(U_0U^\dagger\right)\right|}{2}}, \tag{5}$$

where $U_0$ denotes the matrix representation of the braidword, $U$ represents the target one-qubit gate, the dagger symbol $\dagger$ indicates the conjugate transpose of $U$, and $Tr$ denotes the trace of $U_0U^\dagger$. For convenience, we denote the global phase-invariant distance $d\left(U_0,U\right)$ simply by $d$

The Solovay-Kitaev algorithm (SKA) [31] is a fundamental method in quantum computation for efficiently approximating an arbitrary target gate with a finite universal gate set. The conventional SKA has a significant limitation: its 0-level approximation relies on BF search. When the base length $l_0$ becomes too large, the exponentially growing number of possible sequences makes BF search computationally infeasible due to prohibitively high time costs. Innovatively, Emil Génetay Johansen and Tapio Simula proposed replacing the BF search in the traditional SKA with MC simulations [34]. This modification removes the constraint on $l_0$ size, thereby enabling the construction of higher-precision approximations of standard one-qubit gates at a lower computational cost and enhancing the overall efficiency of the SKA. The GA simulates the principle of survival of the fittest in nature, where individuals within a population with higher fitness to the environment are retained to form a new generation. This new population is generated from the previous generation through hybridization and typically exhibits higher environmental fitness. By iterating this process, individuals with progressively greater adaptability to the environment are obtained [36]. In our previous studies, We employed the GA to replace MC simulations for obtaining the 0-level approximation required by the SKA yielded superior results [9,16,35]. Technical details of the GA-enhanced SKA are provided in SUPPLEMENTAL SECTION II.

Building on our previous work [35], where a GA-enhanced SKA was developed for Fibonacci anyons to construct single-qubit gates, we now extend this approach to the non-semisimple Ising anyon model in order to build a full universal gate set and to numerically demonstrate its universality. To this end, we consider both high-fidelity single-qubit gates constructed from single-qubit EBMs and the CNOT gate constructed from two-qubit EBMs. The non-semisimple Ising anyon model introduces a distinctive dependence of the two-qubit leakage error on the parameter $\alpha$, which prevents a straightforward transfer of the previous method. We therefore implement several key algorithmic adaptations: (i) for two-qubit gate synthesis, the leakage error is incorporated as an explicit constraint within the GA; (ii) for single-qubit gate synthesis, the length of the 0-level approximation is optimized so that the 3-level approximation satisfies the fault-tolerance threshold; (iii) the parameter $\alpha$ is scanned across the interval [2.001, 2.999] with a step size of 0.001, from which we identify a suitable range [2.001, 2.999] that supports universal quantum computation. These refinements allow the GA-based method to effectively handle the unique features of non-semisimple Ising anyons. (For simplicity, the four single-qubit EBMs are labeled as {A, B, C, D} and the eight two-qubit EBMs are labeled as {A, B, C, D, E, F, G, H}.) The optimization analysis of the GA parameters is presented in SUPPLEMENTAL SECTION III**.**

## 3. Numerical Results and Gate Compilation Performance

### 3.1 Two-Qubit Entangling Gate and Leakage Error Analysis

In TQC, entangled gates can be directly achieved via few braiding operations for conventional Ising anyons [24], whereas the Fibonacci anyon model relies on a controlled injection method based on its unique fusion rules [10]. Phillip C. Burke compiled a series of low-leakage-error braidwords approximating the local equivalence class [CNOT] using the two-qubit EBMs of the Fibonacci anyon model [11], providing a novel numerical approach for two-qubit gate compilation. Makhlin first introduced three real parameters, known as local invariants to fully characterize a two-qubit entangled gate [37]. Zhang et al. further incorporated these local invariants into the SU(4) Cartan decomposition, introducing a geometric framework into the study of two-qubit gates and enabling an intuitive visual representation [38]. M. M. Muller et al. demonstrated that optimizing over an entire local equivalence class relaxes control constraints and enhances both flexibility and success rates in gate compilation [39]. Two matrices are considered approximate within a local equivalence class if they can be interconverted via one-qubit operations. A recent study, employing Mixed-Integer Programming, has constructed local equivalence classes of two-qubit entangling gates using the two-qubit EBMs of non-semisimple Ising anyons [40]. In this work, we use the two-qubit EBMs of the non-semisimple Ising anyon model to compile the local equivalence class [CNOT].

Let $B$ denote a braidword formed by composing six-dimensional EBMs. It admits a direct-sum decomposition $B = A \oplus M$, where $A$ represents the four-dimensional computational subspace and $M$ corresponds to the two-dimensional non-computational subspace. The local equivalence classes $g_i(A)$ (i=1, 2, 3) of matrix $A$ are obtained using established method provided in reference [11]. The distance between $A$ and the local equivalence class of the standard CNOT gate is quantified by:

$$d^{\mathrm{CNOT}}(A) = \sum_{i=1}^{3} \Delta g_i^2, \Delta g_i = \left| g_i(A) - g_i(\mathrm{CNOT}) \right|, \tag{6}$$

where $g_1(\mathrm{CNOT}) = 0$, $g_2(\mathrm{CNOT}) = 0$, $g_3(\mathrm{CNOT}) = 1$. The unitarity of A is evaluated by:

$$d^U = Tr(\sqrt{a^\dagger a}), a = A^\dagger A - \mathrm{I}, \tag{7}$$

where I is the four-dimensional identity matrix.

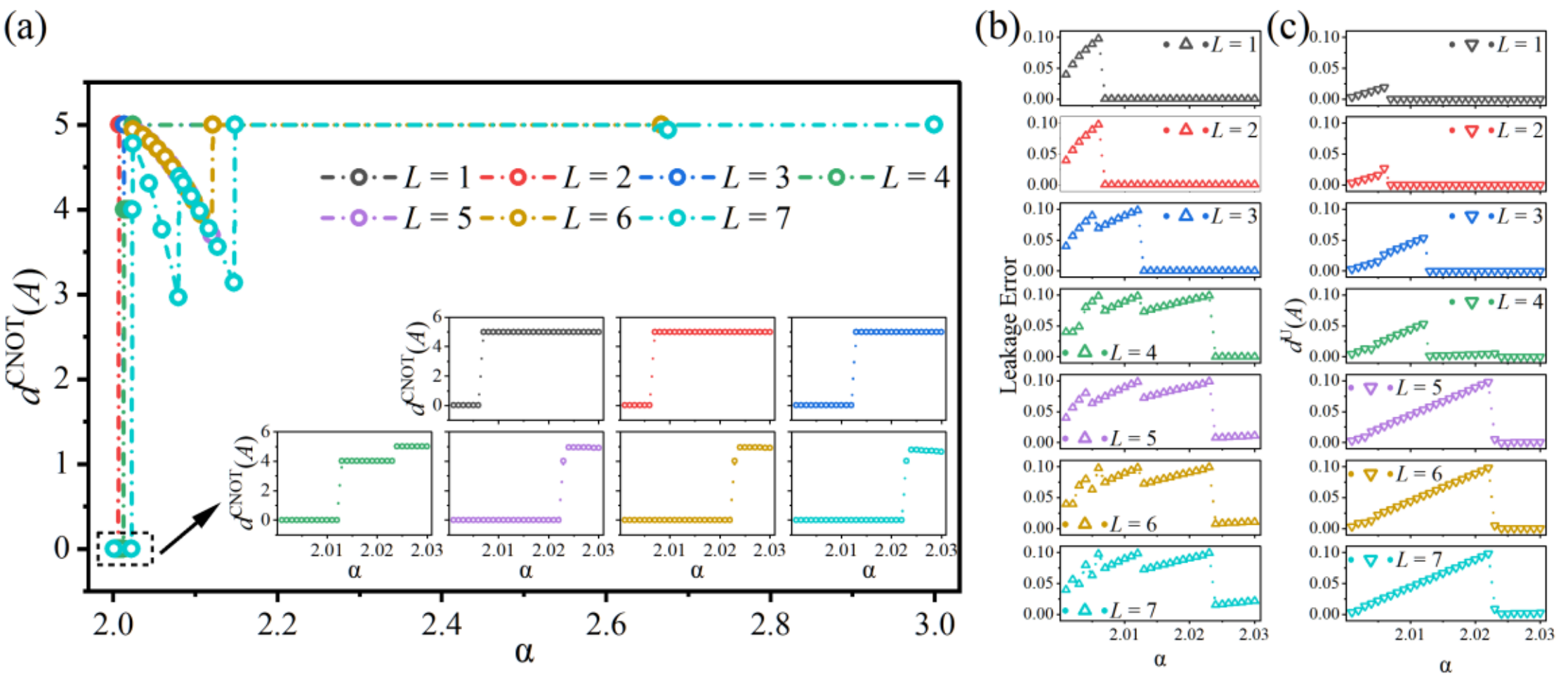

Fig. 2: (a) Computational results for approximating the local equivalence class [CNOT] compiled from EBMs at different α values, subject to unitary measurement of $A$ $d^{\mathrm{U}} < 0.1$, and the leakage error < 0.1. Results from BF search with braid lengths $1 \le L \le 7$. (b) The leakage errors within range α ∈ [2.001, 2.030]. (c) The values of $d^{\mathrm{U}}(A)$ within range α ∈ [2.001, 2.030].

Note that although the operations $\left(b_1^{(3)}\right)^2$ and $\left(b_1^{(5)}\right)^2$ correspond to two physical braiding actions, they are treated as a single unit of braid length in this context. Fig. 2(a) presents result of approximating local equivalence class [CNOT] using braidwords of lengths $L$=1~7 based on EBMs with α varying from 2.001 to 2.999 in increments of 0.001 to determine the specific forms of the EBMs. To investigate the approximation of the local equivalence class [CNOT] with longer braidwords, the distance $d$ for one-qubit (Eq.(5)) is replaced by $d^{\mathrm{CNOT}}(A)$ (Eq.(6), and the one-qubit EBMs {A, B, C, D} are correspondingly replaced by two-qubit EBMs{A, B, C, D, E, F, G, H} in the GA. We constrained the unitary measure $d^{\mathrm{U}}(A) < 0.1$and leakage error < 0.1. Here, leakage error is defined as the magnitude of matrix elements outside the computational and non-computational subspaces for the two-qubit braidword. According to Fig. 2(a), only within a narrow range of α close to 2 in the interval [2.001, 2.999] do the corresponding $d$ values approach 0. With α restricted to [2.001, 2.030] as the length $L$ increases from 1 to 5, the range of α yielding minimal $d$ values gradually expands and then stabilizes at $\alpha \in [2.001, 2.022]$. This trend corresponds to the gradual deterioration of $A$'s unitarity in Fig. 2(c), approaching but not exceeding the upper bound of 0.1. The leakage error in Fig. 2(b) is defined as the maximum value in the non-diagonal part (all 16 matrix elements) of matrix $B$, excluding those of the four-dimensional computational and the two-dimensional non-computational matrix, and all 16 matrix elements are constrained to be below 0.1.

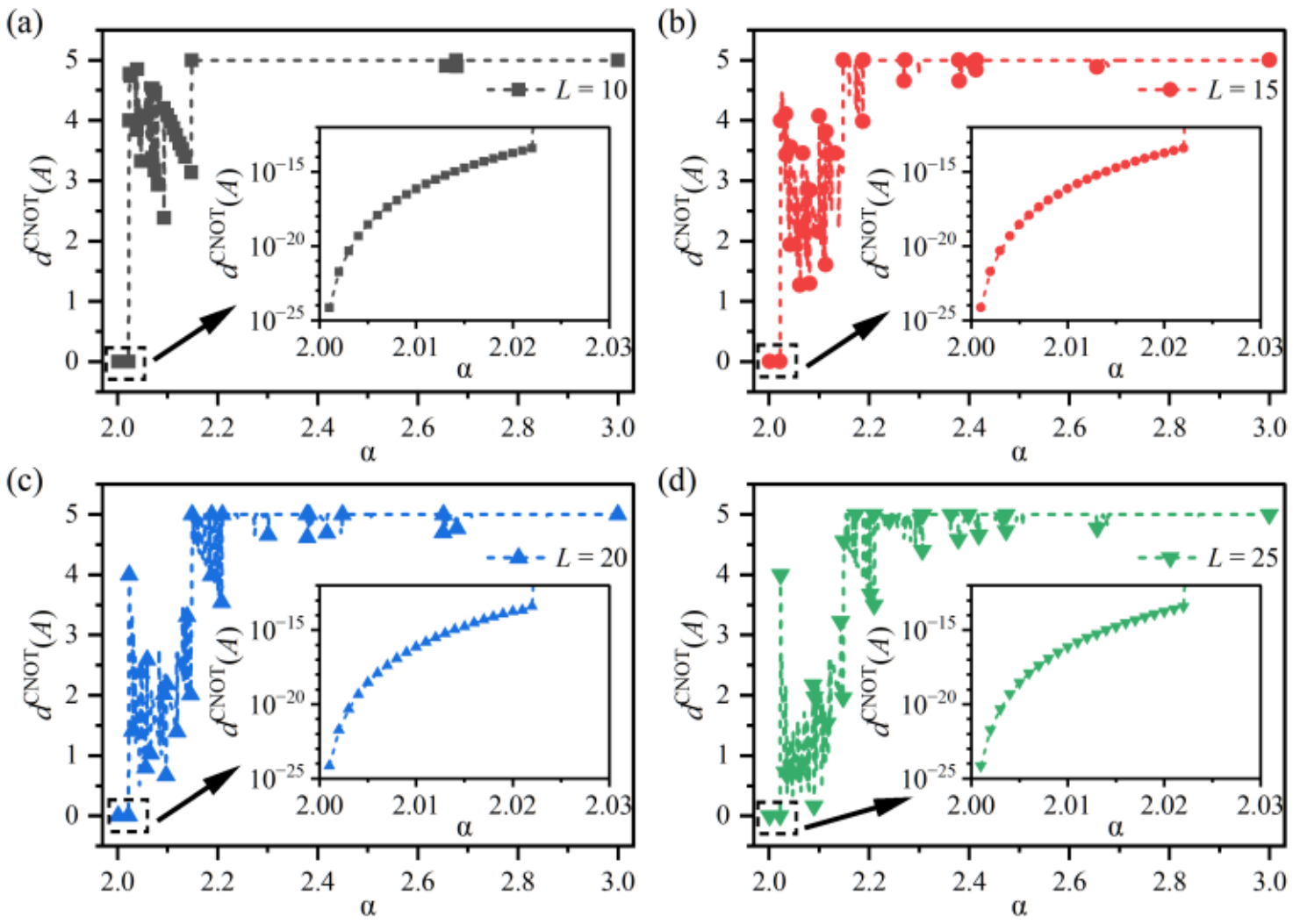

Fig. 3: Computational results for approximating the local equivalence class [CNOT] compiled from EBMs at different α values, subject to unitary measurement of $A$ $d^{\mathrm{U}} < 0.1$, and the leakage error < 0.1. Results from GA with braid lengths $L$ = (a) 10, (b) 15, (c) 20, (d) 25.

Fig. 3 shows results obtained via GA using braidwords of lengths 10, 15, 20, and 25. Although GA overcomes BF search limitations by increasing $L$, it does not significantly extend the α range that can naturally approximate [CNOT]. The α range capable of approximating [CNOT] with $d^{\mathrm{CNOT}}(A) < 10^{-13}$ remains [2.001, 2.022]. While GA reduces $d$ values for $\alpha \in [2.023, 2.999]$ compared to BF search, these reductions are insufficient for high-fidelity requirements and are therefore not considered.

Progressively tightening the leakage error threshold yields analogous results: after $L$ increases from 1 to 5, the range of α naturally approximates [CNOT] reaches its maximum. Further increasing $L$ via GA does not expand this range. A stricter leakage error threshold leads to a narrower viable α-range. An exception occurs at a leakage error of 0.09, where GA extension expands the α-range from [2.001, 2.019] to [2.001, 2.020], despite $d$ at α = 2.020 ($\sim 10^{-9}$) remaining orders of magnitude larger than at α = 2.019 ($\sim 10^{-13}$). The relationship between leakage error and viable α-range is summarized in Table I.

**Table** I. The range of α that can naturally approximate the local equivalence class [CNOT], under the corresponding leakage error constraints.

| The range of α | Leakage errors |
|---|---|
| None | ≤ 0.01 |
| 2.001 | ≤ 0.02 |
| [2.001,2.002] | ≤ 0.03 |
| [2.001,2.004] | ≤ 0.04 |
| [2.001,2.006] | ≤ 0.05 |
| [2.001,2.009] | ≤ 0.06 |
| **[2.001,2.012]** | **≤ 0.07** |
| **[2.001,2.015]** | **≤ 0.08** |
| **[2.001,2.020]** | **≤ 0.09** |
| **[2.001,2.022]** | **≤ 0.10** |

Table II presents braidwords with near-zero $d^{\mathrm{CNOT}}(A)$ at α = 2.012, 2.015, 2.019, 2.020, and 2.022. Braidwords longer than 7 were obtained using the GA, while those shorter than 7 were obtained via BF search; in both cases the $d^{\mathrm{U}}$ was constrained to remain below 0.1.

**Table** II. Braidwords yielding the $d^{\mathrm{CNOT}}(A)$, $d^{\mathrm{U}}$ for the CNOT-gates were obtained via BF search ($L \leq 7$) and GA ($L > 7$).

| | Models | Braidwords | $d^{\mathrm{CNOT}}(A)$ | $d^{\mathrm{U}}$ |
|---|---|---|---|---|
| CNOT-gate | α=2.012 | FHGDB | $3.11345\times10^{-16}$ | 0.05334 |
| | α=2.015 | BHGHB | $1.85601\times10^{-15}$ | 0.06669 |
| | α=2.019 | BGHGB | $1.2302\times10^{-14}$ | 0.08452 |
| | α=2.020 | HEGBCEBCEBEBCDF | $1.21827\times10^{-9}$ | 0.08845 |
| | α=2.022 | FHCDF | $3.97573\times10^{-14}$ | 0.0979 |

### 3.2 Single-Qubit Gate Construction

The specific numerical forms of the EBMs for the non-semisimple Ising anyon model are determined by the parameter α, where α ∈ (2, 3). We systematically varied α from 2.001 to 2.999 in increments of 0.001 to generate a comprehensive set of EBMs. Using these EBMs, BF search was employed to approximate the *H*-/*T*-gates, with the global phase-invariant distance *d* serving as the fidelity metric across braid lengths $L$ ($1 \leq L \leq 13$). For each α and $L$, the minimal distances achieved for the target *H*-/*T*-gates are presented in Fig. 4.

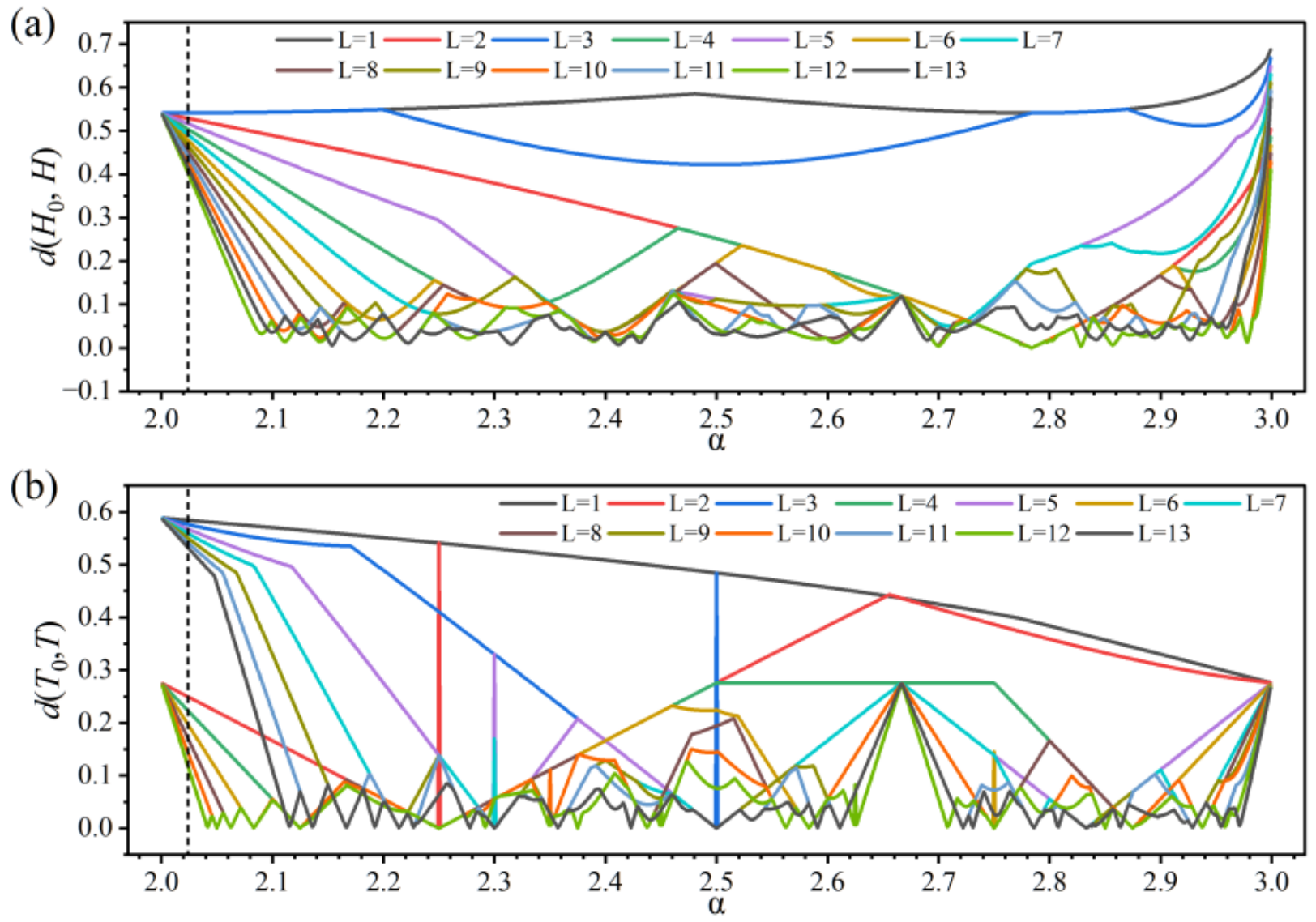

Fig. 4: Minimal $d$ for the (a) $H$-gate and (b) $T$-gate, obtained via BF search using EBMs across $\alpha \in (2, 3)$ with increments of 0.001, with the braid length $L$ increasing from 1 to 13. The dashed line is located at $\alpha = 2.022$.

Low-error approximations of the $H$-gate ($d < 0.1$) were unattainable across the entire α-interval only at braid lengths $L = 1$ and 3. For all other $L$ values, suitable choices of α consistently yielded high-fidelity $H$-gates. Similarly, for the $T$-gate, such low-error approximations were infeasible only at $L = 1$, while all other lengths admitted high-fidelity solutions with appropriate α. Notably, even-length braids generally outperformed odd lengths ones for constructing the $T$-gate.

Our primary objective was to identify a fixed α value for which the corresponding EBMs can be used to construct a universal gate set for the non-semisimple Ising anyon model. As analyzed in Section 3.1, when the unitary measure $A$ of the braidword (Eq. (9)) is set below 0.1, the local equivalence class [CNOT] can be faithfully realized for $\alpha \in [2.001, 2.012]$, provided the leakage error remains below 0.07. Relaxing the leakage error threshold to 0.08, 0.09, and 0.10 extends the feasible range to $\alpha \in [2.001, 2.015]$, $\alpha \in [2.001, 2.020]$, and $\alpha \in [2.001, 2.022]$, respectively. Beyond these α ranges, the two-qubit EBM representations fail to approximate [CNOT] with sufficient accuracy.

However, BF search alone limited to $L \leq 13$ could not produce low-error approximations of the $H$-/$T$-gates for the $\alpha \in [2.001, 2.022]$ (left of the dashed line at $\alpha = 2.022$ in Fig. 4), the achieved minimum distances are $d(H_0, H) = 0.41479$ and $d(T_0, T) = 0.13088$ (for $L = 12$, $\alpha = 2.022$), which are clearly inadequate for fault-tolerant quantum computing. To overcome this length limitation, we employed GA-enhanced SKA to efficiently compile high-fidelity approximations of the $H$- and $T$-gates with $\alpha$ = 2.012, 2.015, 2.220, and 2.022, thereby enabling the construction of a full universal gate set for these four models.

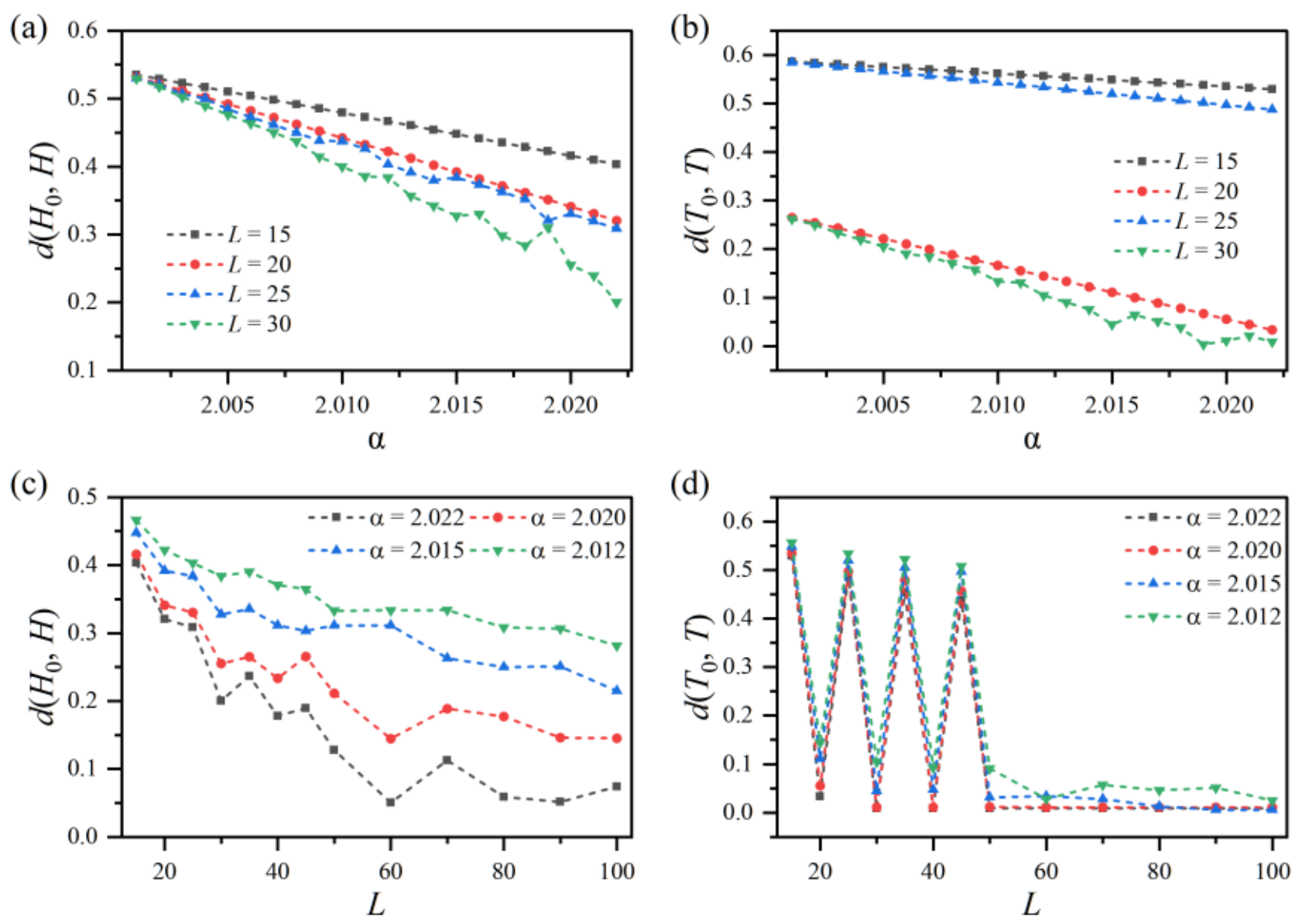

Fig. 5: Converged $d$ for (a) the $H$-gate and (b) the $T$-gate obtained via GA with α ∈ [2.001, 2.022] and braid lengths set to $L$ = 15, 20, 25, 30. Converged $d$ for (c) the $H$-gate and (d) the $T$-gate when braid lengths $L$ ranging from 15 to 100, obtained via GA with α fixed at 2.012, 2.015, 2.20, and 2.022.

Fig. 5(a)/(b) presents the results of constructing the $H$-/$T$-gates via GA using braidwords composed from EBMs with α ∈ [2.001, 2.022] at lengths $L$ = 15, 20, 25, and 30. For both the $H$- and $T$-gates, $d$ exhibits a gradual decrease with increasing α, which is consistent with the BF search results (Fig.4). In the case of the $T$-gate, $L$ = 20 or 30 yields significantly better approximations than $L$ = 15 or 25, consistent with our BF search results indicating superior performance for even-length braids. Furthermore, larger values of α generally correspond to smaller $d$ values at fixed lengths.

Based on following two-qubit results (Table I), we selected four upper boundary α-values (2.012, 2.015, 2.020, and 2.022) corresponding to two-qubit braidword leakage errors below 0.07, 0.08, 0.09, and 0.1, respectively, for further compilation of low-error $H$-/$T$-gates. A prerequisite for recursively constructing high-fidelity gates using SKA is obtaining accurate 0-level approximations. We therefore extended the braid length up to $L$ = 100 via GA using the EBMs at these four α values, with length increments of 5 from 15 to 50, and increments of 10 thereafter up to 100; the results are summarized in Fig. 5(c)/(d). For the $H$-gate, larger α values clearly lead to better approximations, with $d$ decreasing gradually as $L$ increases. For the $T$-gate, even-length braids again consistently outperform odd-length ones. When α is close to 2, the product of an even number of A or C operations approaches the identity matrix I. In this case, the distance $d(T_0, T)$ is close to $d(\mathrm{I}, T)$ (the value approximately 0.27). Continuing to add an even number of A or C operations causes the braidword to gradually approach the standard $T$ gate. When the number of braiding operations is odd, the braidword generally consists of an even number of A or C plus an odd number of B or D. Consequently, $d(T_0, T)$ is close to $d(\mathrm{B}, T)$, which is approximates 0.58 (Fig. 5(b)).

To determine optimal 0-level braids, we selected sequences with the minimal achievable $d$ from the GA, while also considering braid complexity by minimizing length. The chosen base lengths are as follows: for the $H$-gate, $l_0$ = 100 (120), 100, 60, and 80 for α = 2.012 (2.012), 2.015, 2.020, and 2.022, respectively; for the $T$-gate, $l_0$ = 40 (60), 30, 20, and 20 for α = 2.012 (2.012), 2.015, 2.020, and 2.022, respectively. Optimized 0-level braidwords for both gates are summarized in Table III.

**Table** III. 0-level braidwords and $d(U_0, U)$ metrics for $H$-/$T$-gates.

| | Models | Braidwords | $d(U_0, U)$ |
|---|---|---|---|
| $H$-gate | α=2.012 ($l_0$=100) | CDCCCDDADDCCBCBCCCCBAADDADABABADCD<br>CBCBCCCCCBBADCBCBABABCBCCCCCCDCCDAA<br>AADAAABDCBAAAAAABBCCBBADCDAAABA | 0.28154 |
| | α=2.012 ($l_0$=120) | DDDDACBCDABCDCBCCDDADCDDADDDCCDCBB<br>BBBABCBBDADCBCCCCCBBABABAAAAAAADDC<br>BBAAADCDDCBCCBABADCDCBABBBAAAAAAAD<br>CBBADDCBCCCCCBADAD | 0.24227 |
| | α=2.015 ($l_0$=100) | BCDABADABAADCBAAABABCAAAAAAAAAAAAA<br>AADCBCCCBCDADADADDCCBCDDDCDCBBABDC<br>DADCDDCBBCBCBCBCCBCBBCBCCCCBABAB | 0.21498 |
| | α=2.020 ($l_0$=60) | ABAAAAABCBAAAADADADDCCBCCDDDAAAAAA<br>AAADCDAADADAAAAAAAAAADABCB | 0.14447 |
| | α=2.022 ($l_0$=80) | DAAADABCDCBAAAAAAAAAABBCCBADAAAAAD<br>ABCBCBCBCDAAAABDDCDAAADDDADABADADC<br>CBDABBDCCCCB | 0.05880 |
| $T$-gate | α=2.012 ($l_0$=40) | CCCCCDDAAAAADADAAADADAADCDAAAAAAA<br>AADAAAB | 0.09253 |
| | α=2.012 ($l_0$=60) | CCBCDDCBCCCCBCDADABCCCCCCBBAADAADCD<br>DADDCCBBAAAAAAAAAAAAAAABB | 0.02817 |
| | α=2.015 ($l_0$=30) | CCCCCCCCCCCBABCCCCCCCCCCCCCCCC | 0.04442 |
| | α=2.020 ($l_0$=20) | CCCCCCCCCCCCCCCCCCCC | 0.05552 |
| | α=2.022 ($l_0$=20) | CCCCCCCCCCCCCCCCCCCC | 0.03332 |

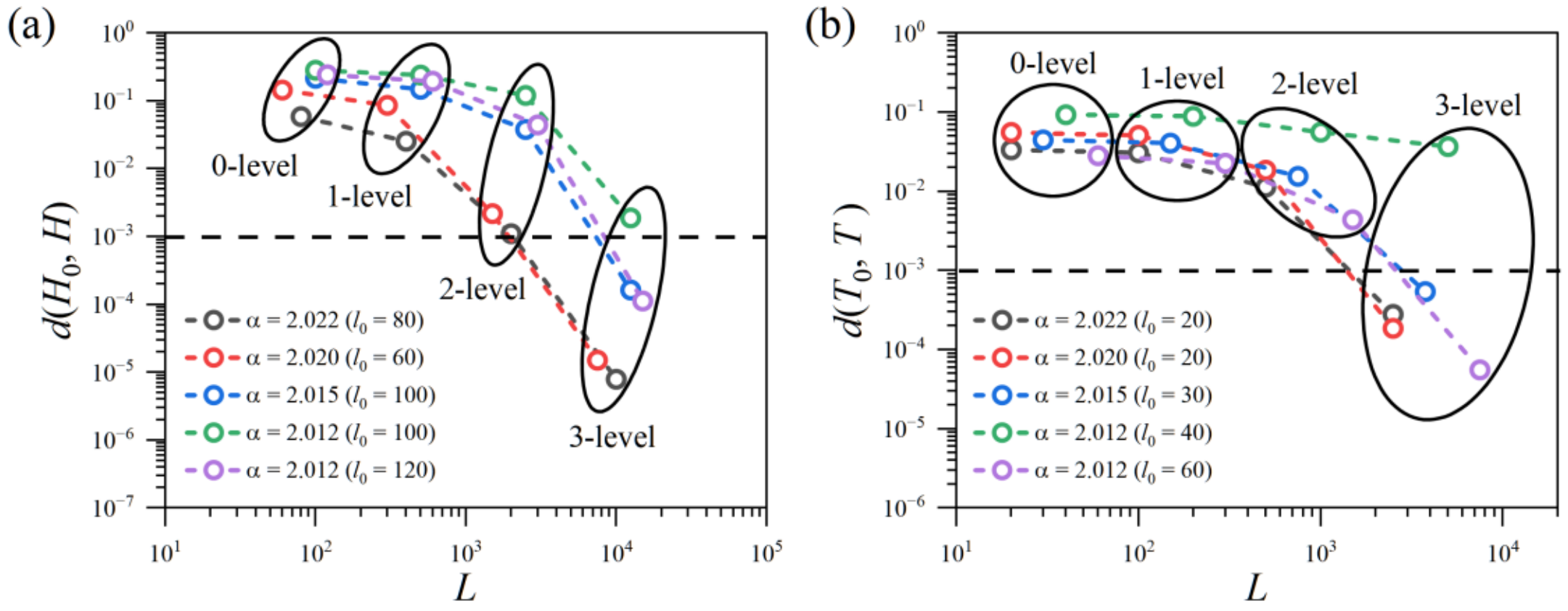

Fig. 6: Construction of the standard (a) *H*-gate and (b) *T*-gate using the GA-enhanced SKA for non-semisimple Ising anyons (α = 2.012, 2.015, 2.20, and 2.022). The circles group data points from different models that achieve the same approximation level of *d*. The dashed line marks $d = 10^{-3}$.

According to the threshold theorem for surface-code quantum error correction [41,42], physical error rates below approximately 1% are tolerable for fault-tolerant quantum computation. However, the gate compilation error *d* represents a systematic approximation error that is independent of, and additive to, physical noise. To prevent the compilation error from dominating the error budget, we adopt the conservative design target of $d < 10^{-3}$. This provides a sufficient safety margin, reserving the bulk of the error budget for physical noise, state preparation, and measurement imperfections. This choice is consistent with standard practice in topological quantum compilation [34].This threshold corresponds to data points below the dashed line in Fig. 6.

Fig. 6(a) and (b) present numerical results for constructing *H*-/*T*-gates using GA-enhanced SKA with EBMs corresponding to α = 2.012, 2.015, 2.020, and 2.022. As shown in Fig. 5(c) and (d), as α decreases, the base length $l_0$ must be increased to achieve sufficiently excellent 0-level approximations. An exception occurs for the *H*-gate at α = 2.022, where $l_0 = 80$ suffices, whereas α = 2.020 requires$l_0 = 60$. This is because only at α = 2.022 with $l_0 = 80$ could a 0-level approximation with $d(H_0, H) < 0.1$ be obtained; for other α values, even extending $l_0$ to 100 yields $d(H_0, H) < 0.1$. We regard this as a favorable starting point, and subsequent calculations confirm that at every approximation level, α = 2.022 gives the smallest $d(H_0, H)$ (corresponding to the lowest data points within each circle in Fig. 6(a) compared to other α values). Using the GA-enhanced SKA, the fault-tolerance requirement ($d < 10^{-3}$) is met at recursion level 3 for these α values (α = 2.012, 2.015, 2.20, and 2.022).

For both *H*- and *T*-gates, results demonstrate that for selected α values and prescribed $l_0$, GA-enhanced SKA successively reduces $d(H_0, H)$ and $d(T_0, T)$. However, as α decreases, the reduction in *d* becomes increasingly limited. In particular, for α = 2.012 ($l_0 = 100$ for *H*, $l_0 = 40$ for *T*), 3-level approximations for both gates yield *d* values above $10^{-3}$, insufficient for fault-tolerant quantum computation. Achieving required precision would demand 4-level approximation, requiring $100 \times 5^4$ and $40 \times 5^4$ braid operations, respectively. This issue was resolved by extending $l_0$ from 100 to 120 for the $H$-gate and from 40 to 60 for the $T$-gate. With α = 2.012 ($l_0$=120, 60), the 3-level approximations give $d(H_0, H)$ and $d(T_0, T)$ below $10^{-3}$, while the corresponding braid counts are reduced to $120 \times 5^3$ and $60 \times 5^3$, thus meeting the fault-tolerance requirement with significantly fewer braiding operations.

Table II with the standard $H$- and $T$-gate compilation results in Fig. 6, constitute a set of universal quantum computation models based on four non-semisimple Ising anyon models. However, constrained by GA performance, constructing a high-precision universal gate set requires controlling leakage error $\in \{0.07, 0.08, 0.09, 0.10\}$. The main challenge is that for α < 2.012, achieving fault-tolerant precision via GA-enhanced SKA is difficult due to excessive base lengths (exceeds 120) and computational complexity of implementing 4-level or higher SKA approximations. Future work should focus on developing more efficient methods to overcome these obstacles and enable high-precision universal gate sets with lower leakage errors.

### 3.3 Benchmarking Gate Compilation Performance for Braiding-Based Anyon Models

The $SU(2)_k$ ($k$>2, $k$≠4) anyon models support universal quantum computation through braiding alone [7]. In these models, 3 or 4 anyons of 1/2-topological spin encode a single-qubit [43]. According to the fusion rule $\frac{1}{2}\times\frac{1}{2}=0+1$, two 1/2-topological spin anyons can fuse either to the vacuum or to a 1-topological spin anyon; this twofold fusion outcome naturally provides the two basis states of a qubit. Arbitrary SU(2) operations on the encoded qubit can be generated by repeated application of single- or double-braiding among these anyons [8]. Six 1/2-topological spin anyons encode two qubits and yield a 5-dimensional EBMs, comprising a 4-dimensional computational subspace and a 1-dimensional non-computational subspace [44]; standard two-qubit gates can be constructed from these EBMs [11]. The EBMs of these $SU(2)_k$ models obey the Artin braid group relations [45]:

$$\begin{aligned} b_i b_j &= b_j b_i \qquad \text{for } |i-j| \geq 2, \\ b_i b_{i+1} b_i &= b_{i+1} b_i b_{i+1}, \end{aligned} \tag{8}$$

where $b_i$ denotes the braiding of the $i$-th and ($i$+1)-th anyons.

The non-semisimple Ising anyon model extends the $SU(2)_2$ theory by introducing a *neglecton* α, which renders the model universal by braiding alone. In this model, α remains stationary throughout the braiding process (braiding the α with a σ anyon requires two consecutive exchanges, which is topologically equivalent to a full winding of σ around α. The braiding operations are therefore a mixture of single- and double - braiding. A single-qubit is formed by one α and two σ anyons, while two additional σ anyons are required to encode two qubits. The resulting two-qubit EBMs is 6-dimensional, with a 4-dimensional computational subspace and a 2-dimensional non-computational subspace dictated by the fusion rules Eq. (1), and these EBMs satisfies the affine braid group relation [26,29]:

$$b_1^2 b_2 b_1^2 b_2 = b_2 b_1^2 b_2 b_1^2. \tag{9}$$

To evaluate the performance of gate compilation via braiding operations for non-semisimple Ising anyons, we compare it with the SU(2)$_k$ anyon models ($k$ = 3, 5, 6, 7, the precise forms of EBMs of these four anyon models are derived from [9,10,44]). For the non-semisimple Ising model, the parameter α is chosen as 2.022, as it demonstrates an advantage in constructing $H$ and $T$ gates over other values of α, as shown in Figs. 6(a) and 6(b).

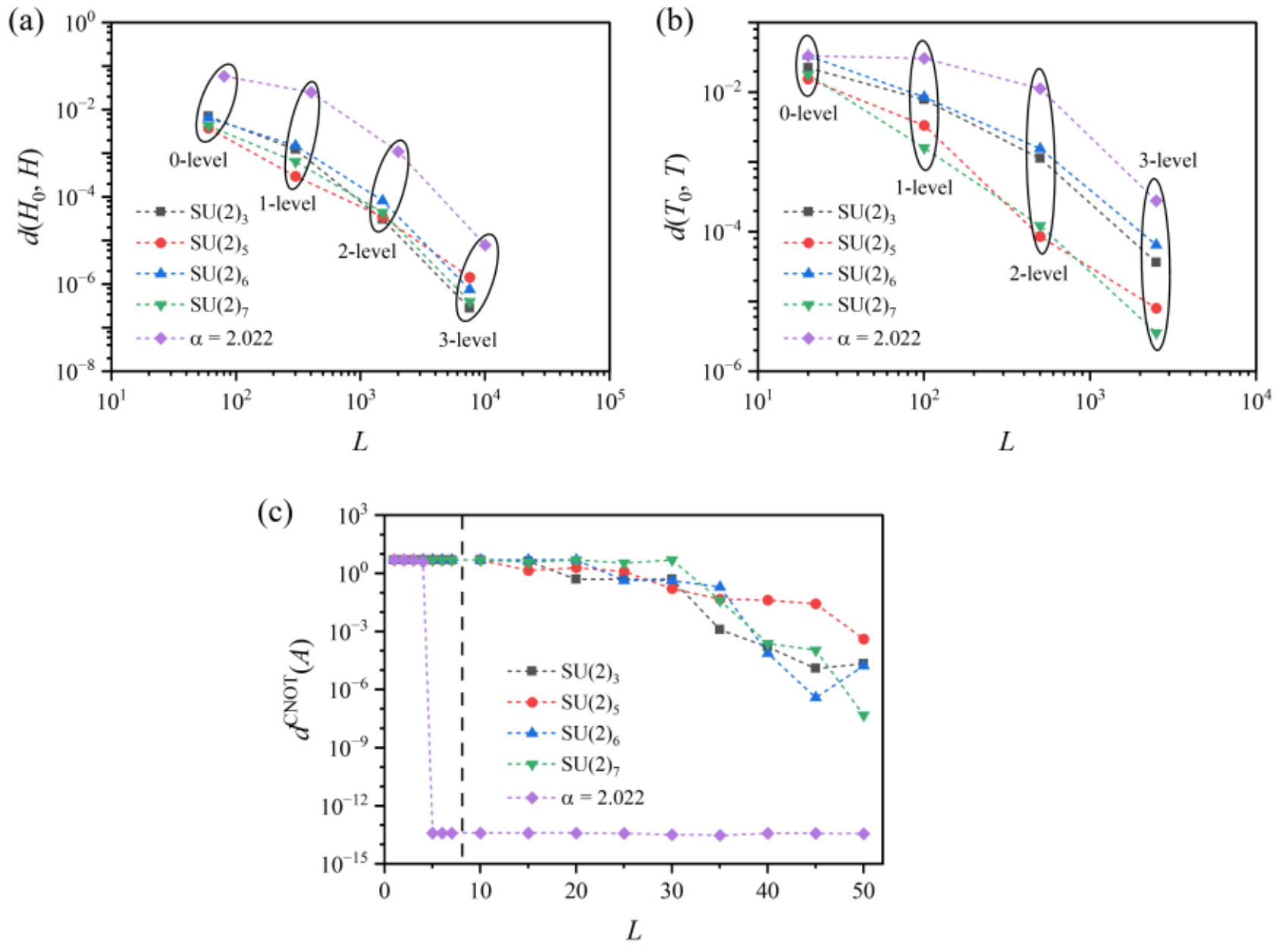

Fig. 7: Construction of the standard (a) $H$-gate and (b) $T$-gate using one-qubit EBMs of SU(2)$_k$ ($k$ = 3, 5, 6, 7) anyon models and non-semisimple Ising anyon models (α = 2.022). (c) Results of approximating the local equivalence class [CNOT] using two-qubit EBMs of SU(2)$_k$ ($k$ = 3, 5, 6, 7) anyon models and non-semisimple Ising anyon models (α = 2.022), the leakage error < 0.1. The boundary between BF search and GA results is indicated by dashed line.

Figs. 7(a) and 7(b) present the results of compiling the standard $H$ and $T$ gates using the SU(2)$_k$ ($k$ = 3, 5, 6, 7) anyon models and the non-semisimple Ising anyon model (α = 2.022) via the GA-enhanced SKA. The GA-enhanced SKA enables each model to exponentially reduce the $d$ between the braidword and the target gate. For the $H$ gate, the $d$ achieved by the SU(2)$_k$ models at each level are comparable, while the non-semisimple Ising model (α = 2.022) performs roughly one order of magnitude worse. Specifically, at 3-level, the $d$ is about $10^{-6}$ for the SU(2)$_k$ models, whereas for the non-semisimple Ising model (α = 2.022) it remains around $10^{-5}$. For the $T$ gate, SU(2)$_5$ and SU(2)$_7$ outperform SU(2)$_3$ and SU(2)$_6$, which in turn surpass the non-semisimple Ising model (α = 2.022).

Fig. 7(c) compares the performance in approximating the local equivalence class [CNOT]. Here, the leakage errors and the $d^{\mathrm{U}}(A)$ are constrained to below 0.1. For the non-semisimple Ising model (α = 2.022), $d^{\mathrm{CNOT}}(A)$ drops sharply to $10^{-14}$ once the $L$ reaches 5, but further increasing the $L$ does not reduce $d$ further. In contrast, for the $SU(2)_k$ models, a noticeable decrease in $d^{CNOT}(A)$ occurs only when the $L$ reaches 30 and cannot obtain the precision of non-semisimple Ising model even $L$=50 ($d^{CNOT}(A)=10^{-7}\sim10^{-3}$ within $L$=30~50). Notably, although the precision of non-semisimple Ising anyon model is slightly less than the $SU(2)_k$ models in compiling standard single-qubit gates (Fig. 7(a)(b)), it holds a significant advantage in compiling the entangling CNOT gate (Fig. 7(c)). It requires only a few braiding operations to approximate the local equivalence class [CNOT] with an exceptionally small $d^{CNOT}(A)$.

**4 Conclusion and Future Prospects**

In this work, we investigate the computational universality of a recently proposed non-semisimple Ising anyon model. This model extends the conventional Ising anyon theory within the framework of non-semisimple topological quantum field theory. The extension introduces a new type of quasiparticle, termed the *neglecton*, parameterized by a continuous real parameter $\alpha \in (2,3)$. Compared with the standard $SU(2)_2$ Ising anyons, the fusion rules become substantially richer, and new $F$-symbols and $R$-symbols emerge, which fundamentally modify the braiding properties of the model.

Our main result is an explicit numerical demonstration that universal quantum computation can be achieved in this model through braiding operations alone, both arbitrary single-qubit rotations and a two-qubit entangling operation can be realized. This overcomes a well-known fundamental limitation of the standard Ising anyon model, where braiding alone generates only the Clifford group and cannot produce the crucial $\pi/8$ phase gate ($T$ gate). By systematically constructing braiding sequences for the $H$ gate, the $T$ gate, and the CNOT gate, we show that a complete universal gate set $\{H, T, \text{CNOT}\}$ can be realized for models with $\alpha \in [2.001, 2.022]$. A particularly important finding is that the CNOT gate can be approximated with high precision using only a few numbers of braids; its approximation fidelity is fundamentally limited only by the leakage of quantum information from the computational subspace into non-computational fusion channels. This leakage error corresponds to transitions into physically distinct anyonic charge sectors and is kept below 0.07–0.10, depending on the chosen value of α. We employ an GA-enhanced SKA to efficiently search the exponentially growing space of braiding sequences. This approach enables us to find high-fidelity single-qubit gate approximations that meet fault-tolerance requirements (error below $10^{-3}$) while maintaining a manageable braid length. We also perform a systematic comparison with the well-studied Fibonacci anyon model ($SU(2)_3$) and other $SU(2)_k$ models. The results indicate that, although the non-semisimple Ising model requires longer braid sequences for single-qubit gates, it exhibits a distinct advantage in two-qubit gate compilation, it can realize a nearly exact CNOT gate approximation with far fewer braiding operations, this highlights the unique value of non-semisimple Ising anyons as a physical platform for TQC. Future work should focus on developing more efficient compilation methods beyond the GA to further suppress the leakage error of entangling gates and to extend the range of α over which high-fidelity universal gate sets can be constructed.

In summary, this work establishes a viable and novel pathway toward universal quantum computation using non-semisimple Ising anyons, providing a robust numerical foundation for future theoretical and experimental explorations in TQC. Future work will also focus on implementing elementary quantum algorithms, such as the Deutsch algorithm, directly using the braid sequences developed in this work, in order to demonstrate the practical utility of non-semisimple Ising anyons for quantum computation.

**Funding** This work is supported by the National Natural Science Foundation of China (Grant Nos. 12374046, 11204261), College of Physics and Optoelectronic Engineering training program, a Key Project of the Education Department of Hunan Province (Grant No. 19A471), Natural Science Foundation of Hunan Province (Grant No. 2018JJ2381), Shanghai Science and Technology Innovation Action Plan (Grant No. 24LZ1400800), Education Department of Hunan Province (Grant No. 24C0316).

**Data Availability Statement** The datasets generated during and/or analyzed during the current study are available from the corresponding author on reasonable request

**Declarations**

**Conflict of interest** No potential conflict of interest was reported by the authors. All authors of this manuscript have read and approved the final version submitted, and contents of this manuscript have not been copyrighted or published previously and are not under consideration for publication elsewhere.

## References

[1] A. W. Harrow and A. Montanaro. Quantum computational supremacy, Nature. **549**, 203 (2017). https://doi.org/10.1038/nature23458.

[2] A. Y. Kitaev. Fault-tolerant quantum computation by anyons, Annals of Physics. **303**, 2 (2003). https://doi.org/10.1016/S0003-4916(02)00018-0.

[3] B. Field and T. Simula. Introduction to topological quantum computation with non-Abelian anyons, Quantum Science and Technology. **3**, 045004 (2018). https://doi.org/10.1088/2058-9565/aacad2.

[4] J. K. Pachos, *Introduction to topological quantum computation* (Cambridge University Press, 2012).

[5] P. Bonderson, M. Freedman, and C. Nayak. Measurement-Only Topological Quantum Computation, Physical Review Letters. **101**, 010501 (2008). https://doi.org/10.1103/PhysRevLett.101.010501.

[6] C. Nayak, S. H. Simon, A. Stern, M. Freedman, and S. Das Sarma. Non-Abelian anyons and topological quantum computation, Reviews of Modern Physics. **80**, 1083 (2008). https://doi.org/10.1103/RevModPhys.80.1083.

[7] M. H. Freedman, M. J. Larsen, and Z. Wang. The Two-Eigenvalue Problem and Density¶of Jones Representation of Braid Groups, Communications in Mathematical Physics. **228**, 177 (2002). https://doi.org/10.1007/s002200200636.

[8] A. L. Kaufmann and S. X. Cui. Universal topological quantum computing via double-braiding in SU(2) Witten–Chern–Simons theory, Quantum Information Processing. **24**, 14

(2025). https://doi.org/10.1007/s11128-024-04633-1.
[9] J. Long, Y. Li, J. Zhong, and L. Meng. The construction of a universal quantum gate set for the SU(2)k (k=5,6,7) anyon models via genetic optimized algorithm, Physics Letters A. **565**, 131142 (2026). https://doi.org/10.1016/j.physleta.2025.131142.
[10] L. Hormozi, G. Zikos, N. E. Bonesteel, and S. H. Simon. Topological quantum compiling, Physical Review B. **75**, 165310 (2007). https://doi.org/10.1103/PhysRevB.75.165310.
[11] P. C. Burke, C. Aravanis, J. Aspman, J. Mareček, and J. Vala. Topological quantum compilation of two-qubit gates, Physical Review A. **110**, 052616 (2024). https://doi.org/10.1103/PhysRevA.110.052616.
[12] A. Tounsi, N. E. Belaloui, M. M. Louamri, A. Benslama, and M. T. Rouabah. Optimized topological quantum compilation of three-qubit controlled gates in the Fibonacci anyon model: A controlled-injection approach, Physical Review A. **110**, 012603 (2024). https://doi.org/10.1103/PhysRevA.110.012603.
[13] H. Xu and J. Taylor. Unified approach to topological quantum computation with anyons: From qubit encoding to Toffoli gate, Physical Review A. **84**, 012332 (2011). https://doi.org/10.1103/PhysRevA.84.012332.
[14] S. X. Cui and Z. Wang. Universal quantum computation with metaplectic anyons, Journal of Mathematical Physics. **56**, (2015). https://doi.org/10.1063/1.4914941.
[15] C. Levaillant, B. Bauer, M. Freedman, Z. Wang, and P. Bonderson. Universal gates via fusion and measurement operations on SU (2) 4 anyons, Physical Review A. **92**, 012301 (2015). https://doi.org/10.1103/PhysRevA.92.012301.
[16] J. Long, J. Zhong, and L. Meng. Topological quantum compilation of metaplectic anyons based on the genetic optimized algorithms, Physical Review A. **112**, 022421 (2025). https://doi.org/10.1103/js85-r2m7.
[17] M. Ilyas, S. Cui, and M. Perkowski. Ternary logic design in topological quantum computing, Journal of Physics A: Mathematical and Theoretical. **55**, 305302 (2022). https://doi.org/10.1088/1751-8121/ac7b55.
[18] S. Mironov and A. Morozov. Entangling gates from cabling of knots, The European Physical Journal C. **85**, 799 (2025). https://doi.org/10.1140/epjc/s10052-025-14492-4.
[19] Y. Kasahara *et al.* Majorana quantization and half-integer thermal quantum Hall effect in a Kitaev spin liquid, Nature. **559**, 227 (2018). https://doi.org/10.1038/s41586-018-0274-0.
[20] D. Wang *et al.* Evidence for Majorana bound states in an iron-based superconductor, Science. **362**, 333 (2018). https://doi.org/10.1126/science.aao1797.
[21] S.-Q. Zhang, J.-S. Hong, Y. Xue, X.-J. Luo, L.-W. Yu, X.-J. Liu, and X. Liu. Ancilla-free scheme of deterministic topological quantum gates for Majorana qubits, Physical Review B. **109**, 165302 (2024). https://doi.org/10.1103/PhysRevB.109.165302.
[22] M. Ezawa. Systematic construction of topological-nontopological hybrid universal quantum gates based on many-body Majorana fermion interactions, Physical Review B. **110**, 045417 (2024). https://doi.org/10.1103/PhysRevB.110.045417.
[23] Y.-M. Zhan, G.-D. Mao, Y.-G. Chen, Y. Yu, and X. Luo. Dissipationless topological quantum computation for Majorana objects in the sparse-dense mixed encoding process, Physical Review A. **110**, 022609 (2024). https://doi.org/10.1103/PhysRevA.110.022609.
[24] Z. Fan and H. de Garis. Braid matrices and quantum gates for Ising anyons topological

quantum computation, The European Physical Journal B. **74**, 419 (2010). https://doi.org/10.1140/epjb/e2010-00087-4.
[25] P. Bonderson, D. J. Clarke, C. Nayak, and K. Shtengel. Implementing Arbitrary Phase Gates with Ising Anyons, Physical Review Letters. **104**, 180505 (2010). https://doi.org/10.1103/PhysRevLett.104.180505.
[26] F. Iulianelli, S. Kim, J. Sussan, and A. D. Lauda. Universal quantum computation using Ising anyons from a non-semisimple topological quantum field theory, Nature Communications. **16**, 6408 (2025). https://doi.org/10.1038/s41467-025-61342-8.
[27] B. W. Reichardt. Systematic distillation of composite Fibonacci anyons using one mobile quasiparticle, Quantum Information & Computation. **12**, 876 (2012). https://dl.acm.org/doi/abs/10.5555/2481580.2481590.
[28] D. Chernyak, A. M. Gainutdinov, and H. Saleur. $$ {U}_{\mathfrak{q}}{\mathfrak{sl}}_2 $$-invariant non-compact boundary conditions for the XXZ spin chain, Journal of High Energy Physics. **2022**, 16 (2022). 10.1007/JHEP11(2022)016.
[29] F. Iulianelli, S. Kim, J. Sussan, and A. D. Lauda. Robust universal braiding with nonsemisimple Ising anyons, Physical Review A. **113**, 042622 (2026). https://doi.org/10.1103/bwpz-4kgt.
[30]Z. Wang, *Topological Quantum Computation* (American Mathematical Society, 2010), Vol. 112.
[31] C. M. Dawson and M. A. Nielsen. The Solovay-Kitaev algorithm, Quantum Information & Computation. **6**, 81 (2006). https://dl.acm.org/doi/abs/10.5555/2011679.2011685.
[32] R. B. McDonald and H. G. Katzgraber. Genetic braid optimization: A heuristic approach to compute quasiparticle braids, Physical Review B. **87**, 054414 (2013). https://doi.org/10.1103/PhysRevB.87.054414.
[33] Y.-H. Zhang, P.-L. Zheng, Y. Zhang, and D.-L. Deng. Topological quantum compiling with reinforcement learning, Physical Review Letters. **125**, 170501 (2020). https://doi.org/10.1103/PhysRevLett.125.170501.
[34] E. Génetay Johansen and T. J. P. Q. Simula. Fibonacci anyons versus Majorana fermions: A Monte Carlo approach to the compilation of braid circuits in SU (2) k anyon models, PRX Quantum. **2**, 010334 (2021). https://doi.org/10.1103/PRXQuantum.2.010334.
[35] J. Long, X. Huang, J. Zhong, and L. Meng. Genetic algorithm enhanced Solovay-Kitaev algorithm for quantum compiling of Fibonacci anyons, Physica Scripta. (2025). https://doi.org/10.1088/1402-4896/ae1d30.
[36] J. H. Holland. Genetic Algorithms, Scientific American. **267**, 66 (1992). https://doi.org/10.1038/scientificamerican0792-66.
[37] Y. Makhlin. Nonlocal properties of two-qubit gates and mixed states, and the optimization of quantum computations, Quantum Information Processing. **1**, 243 (2002). https://doi.org/10.1023/A:1022144002391.
[38] J. Zhang, J. Vala, S. Sastry, and K. B. Whaley. Geometric theory of nonlocal two-qubit operations, Physical Review A. **67**, 042313 (2003). https://doi.org/10.1103/PhysRevA.67.042313.
[39] M. M. Müller, D. M. Reich, M. Murphy, H. Yuan, J. Vala, K. Whaley, T. Calarco, and C.

Koch. Optimizing entangling quantum gates for physical systems, Physical Review A. **84**, 042315 (2011). https://doi.org/10.1103/PhysRevA.84.042315.
[40] P. C. B. Pavel Rytir, Christos Aravanis, Jiri Vala, Jakub Marecek. Topological Quantum Compilation Using Mixed-Integer Programming, https://arxiv.org/abs/2511.09513.
[41] A. G. Fowler, A. M. Stephens, and P. Groszkowski. High-threshold universal quantum computation on the surface code, Physical Review A. **80**, 052312 (2009). https://doi.org/10.1103/PhysRevA.80.052312.
[42] A. G. Fowler, M. Mariantoni, J. M. Martinis, and A. N. Cleland. Surface codes: Towards practical large-scale quantum computation, Physical Review A. **86**, 032324 (2012). https://doi.org/10.1103/PhysRevA.86.032324.
[43] R. Ainsworth and J. K. Slingerland. Topological qubit design and leakage, New Journal of Physics. **13**, 065030 (2011). https://doi.org/10.1088/1367-2630/13/6/065030.
[44] S. X. Cui, K. T. Tian, J. F. Vasquez, Z. Wang, and H. M. Wong. The search for leakage-free entangling Fibonacci braiding gates, Journal of Physics A: Mathematical and Theoretical. **52**, 455301 (2019). https://doi.org/10.1088/1751-8121/ab488e.
[45] E. Artin. Theory of Braids, Annals of Mathematics. **48**, 101 (1947). https://doi.org/10.2307/1969218.